\documentclass[graybox]{svmult}

\usepackage{type1cm}        
\usepackage{makeidx}         
\usepackage{graphicx}        
\usepackage{multicol}        
\usepackage[bottom]{footmisc}

\usepackage{newtxtext}       %
\usepackage[varvw]{newtxmath}       

\usepackage{siunitx}

\usepackage{booktabs}

\usepackage{array}
\newcommand{\PreserveBackslash}[1]{\let\temp=\\#1\let\\=\temp}
\newcolumntype{C}[1]{>{\PreserveBackslash\centering}p{#1}}
\newcolumntype{R}[1]{>{\PreserveBackslash\raggedleft}p{#1}}
\newcolumntype{L}[1]{>{\PreserveBackslash\raggedright}p{#1}}

\makeindex             

\graphicspath{{./figures/}}

\begin{document}

\title*{Convolutional neural networks for automated cellular automaton classification}
\author{Michiel Rollier, Aisling J.~Daly, Jan M.~Baetens}
\institute{Michiel Rollier \at Faculty of Bioscience Engineering, Ghent University, Coupure Links 653, 9000 Gent, Belgium, \email{michiel.rollier@ugent.be}}
%
%
\maketitle

\abstract{The emergent dynamics in spacetime diagrams of cellular automata (CAs)
is often organised by means of a number of behavioural classes.
Whilst classification of elementary CAs is feasible and well-studied,
non-elementary CAs are generally too diverse and numerous to exhaustively classify manually.
In this chapter we treat the spacetime diagram as a digital image,
and implement simple computer vision techniques to perform an automated classification of elementary cellular automata
into the five Li-Packard classes.
In particular, we present a supervised learning task to a convolutional neural network,
in such a way that it may be generalised to non-elementary CAs.
If we want to do so,
we must divert the algorithm's focus away from the underlying `microscopic' local updates.
We first show that previously developed deep learning approaches have in fact been trained to identify the local update rule,
rather than directly focus on the mesoscopic patterns that are associated with the particular behavioural classes.
By means of a well-argued neural network design,
as well as a number of data augmentation techniques,
we then present a convolutional neural network that performs nearly perfectly at identifying the behavioural class,
without necessarily first identifying the underlying microscopic dynamics.}

\section{Introduction}\label{sec:introduction}

\begin{figure}
    \sidecaption
    \includegraphics[height=.2\linewidth]{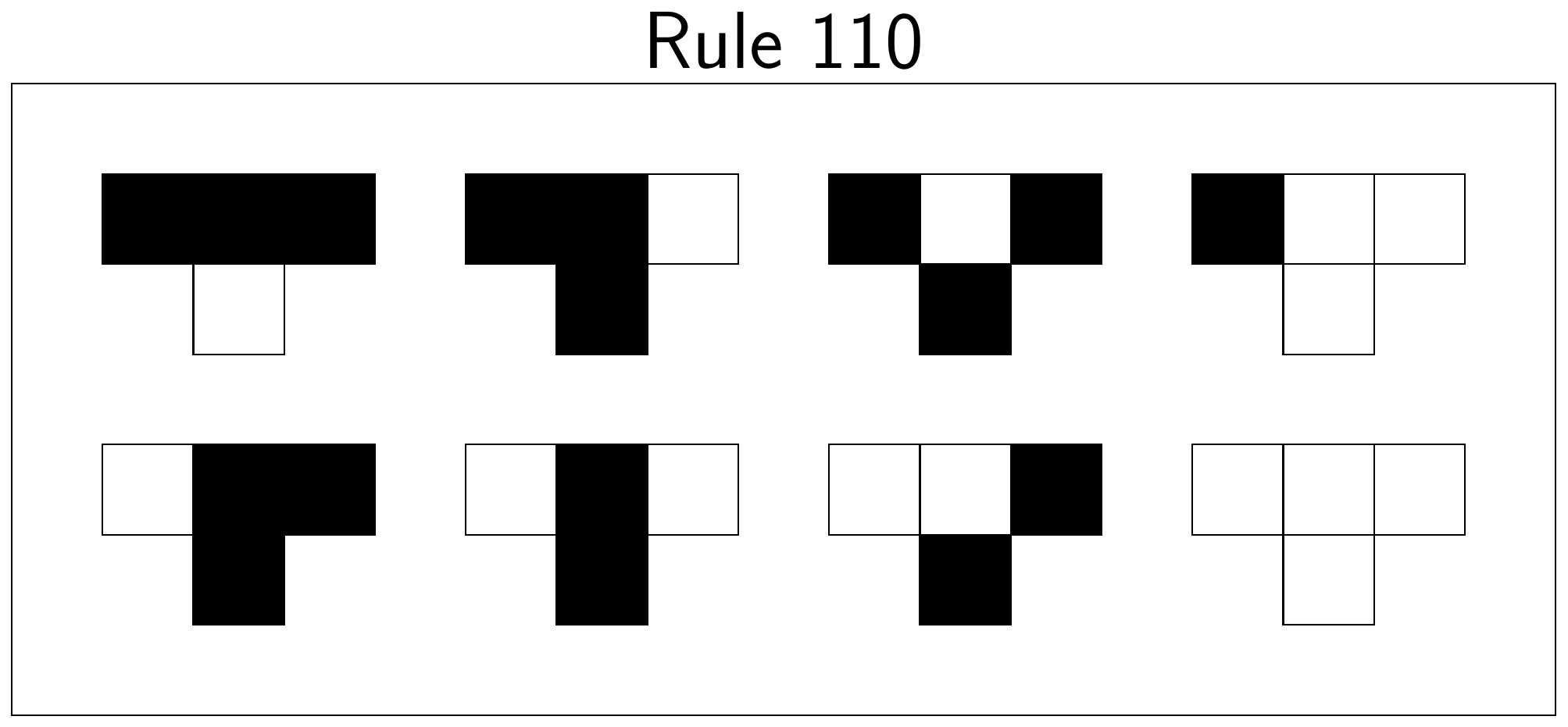}
    \includegraphics[height=.2\linewidth]{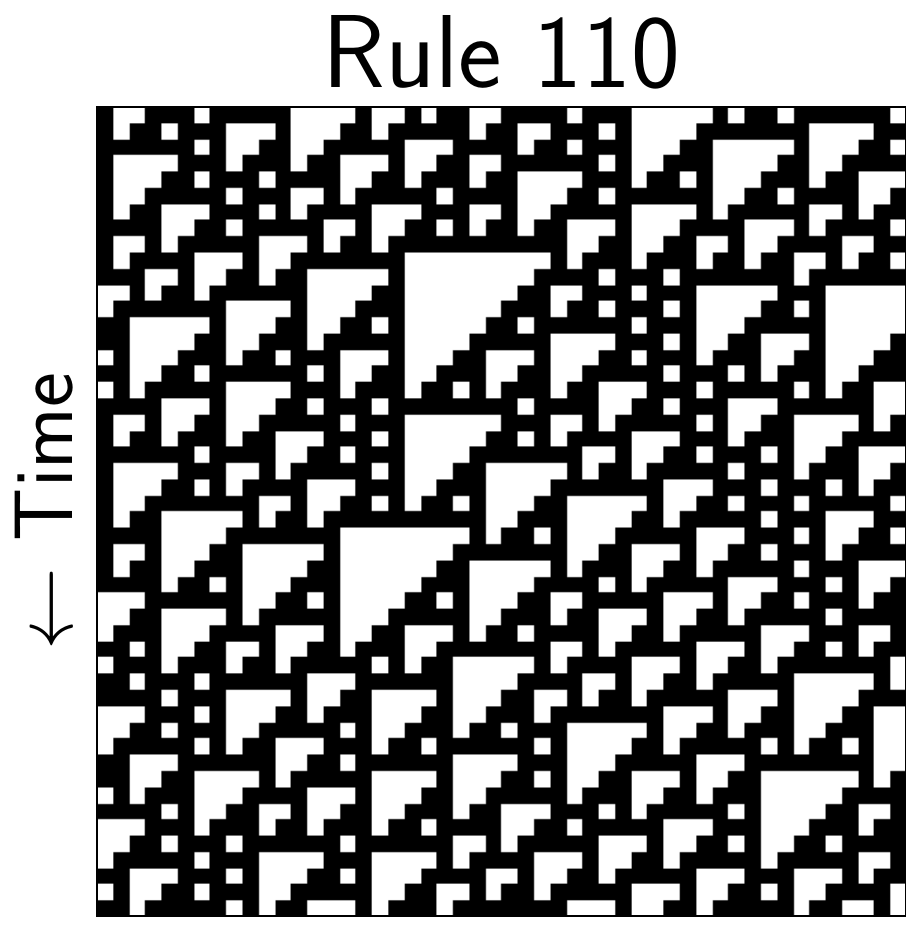}
    \caption{An ECA is governed by a Boolean function with three inputs and one output,
    summarised in the rule table (left).
    An initial (Boolean) configuration is evolved in discrete timesteps
    and generates a spacetime diagram (right).}
    \label{fig:rule110_ruletable_diagram}
\end{figure}

The emergent behaviour of elementary cellular automata (ECAs),
upon evolution from an initial configuration following a local update rule,
is typically displayed in a so-called spacetime diagram (Fig.~\ref{fig:rule110_ruletable_diagram}).
Patterns in this diagram can generally be identified as belonging to a particular behavioural class.
Many ECA diagrams display some repetitive emergent behaviour, for example,
allowing for a collective characterisation as `periodic' ECAs.
In this chapter, we develop and evaluate a neural network (NN)
that is capable of establishing such a classification automatically,
based on rudimentary computer vision techniques.

We design and train the network in such a way
that it achieves excellent performance in identifying the behavioural class,
without necessarily inferring the (elementary) local update rule.
This achieves two goals.
First, we essentially train the network to focus on mesoscopic structures,
rather than patterns at the pixel level.
Such structures are a hallmark of emergent complexity (e.g.~gliders in the Game of Life).
Complexity, in turn, is a property of non-trivial CAs of the highest interest
to mathematicians, computer scientists, and mathematical modellers~\cite{wolfram1994complexity}.
Second, if the network can classify the CA without the need to know
the governing elementary rule, it may also be suitable for identifying interesting classes
in non-elementary CAs.\\

Combining both goals, a sophisticatedly developed NN
would be able to automatically identify complex emergent behaviour
in a set of spacetime diagrams from non-elementary CAs
such as non-uniform or multi-state CAs
(see Rollier et al.~\cite{rollier2024comprehensive} for an overview of CA families).
The number of possible variations in non-elementary CAs quickly becomes overwhelmingly large.
Human classification-by-inspection is therefore liable to missing certain patterns,
and is simply too large of a task to perform manually.
An automatic classification tool constructed from the principles of computer vision
on the other hand, is objective, scrutinous, and fast.
We present the basics ingredients of such a tool in this chapter,
explore a number of strategies for our objective,
and quantitatively compare our results.\\

This chapter's first Section contains an overview of the CA classification problem,
and a very brief introduction to image classification using deep learning.
We conclude the introduction with a clear definition of the available data and the research objective.

\subsection{Behavioural classification of cellular automata}\label{subsec:classification}

Systematic behavioural classification is generally an efficient means to expose
underlying structure in a heterogeneous set of phenomena.
This is certainly also the case for ECA classification,
for which it has been known for a long time that there are strong correlations between
e.g.~the Langton parameter and the resulting behavioural types~\cite{langton1990computation}.
There is no a priori `correct' way of classifying the various ECAs,
but a number of useful approaches have been proposed,
the most familiar (and earliest) being the phenomenological four-fold classification of Wolfram~\cite{wolfram1994complexity}.
Martínez~\cite{martinez2013classification} proposes a different classification,
and in addition reviews 15 existing classifications (see Tab.~\ref{tab:martinez-classes}).
All classifications are based on some properties
of the ECA, an overview of which was compiled by Vispoel et al.~\cite{vispoel2022progress}.
Generally, one distinguishes properties associated with the local update rule (the CA's genotype)
from properties associated with the spacetime diagram (the phenotype).
Our approach will focus on the phenotype.\\

\begin{table}
    \caption{17 different approaches to classifying ECAs.
    We refer to Martínez~\cite{martinez2013classification} for details
    and references to the original work.}
    \begin{tabular}{L{.3\linewidth}L{.57\linewidth}R{0.1\linewidth}}
        \hline\noalign{\smallskip}
        \textbf{Classification} & \textbf{Classes} & \textbf{\# classes} \\
        \noalign{\smallskip}\svhline\noalign{\smallskip}
        Wolfram & Uniform, periodic, chaotic, complex & 4 \\
        ECA with memory & Strong, moderate, weak & 3 \\
        Li-Packard & Null, fixed point, periodic, locally chaotic, chaotic & 5 \\
        Wuensche & Symmetric, semi-symmetric, full symmetric & 3 \\
        Index complexity & Red, blue, green & 3 \\
        Density parameter & P, H, C & 3 \\
        Communication complexity & Bounded, linear, other & 3 \\
        Topological & Period-1, period-2, period-3, Bernouilli $\sigma_t$-shift,
            complex Bernouilli shift, hyper Bernouilli shift & 6 \\
        Power spectrum & Low power density, broad-band noise, power law & 3\\
        Morphological diversity & Chaotic, complex, periodic,
            two-cycle, fixed point, null & 6 \\
        (Non-)distributive lattices & Classes 1, 2, 3, and 4 & 4\\
        Topological dynamics & Equicontinuous, almost-equicontinuous,
            sensitive, sensitive positively expansive & 4 \\
        Expressivity analysis & 0, periodic patterns, complex,
            Sierpinski patterns, finite growth & 5 \\
        Normalised compression & $C_{1,2}, C_{3,4}$ & 2 \\
        Surface dynamics & Types A, B, and C & 3 \\
        Spectral & DE/SFC, DE/SFC SFC, EB, S & 4 \\
        Bijective and surjective & Bijective, surjective & 2 \\
        Creativity & creative, schizophrenic, autistic savants,
            severely autistic & 4 \\
        \noalign{\smallskip}\hline\noalign{\smallskip}
    \end{tabular}
    \label{tab:martinez-classes}
\end{table}

In this chapter, we adopt the Li-Packard (LP) classification of ECAs
(see Li and Packard (1990)~\cite{li1990structure} and Tab.~\ref{tab:Li-Packard-classes}),
for three reasons.
First, because it is a common and well-studied classification,
which facilitates comparison of our results with those found in the literature.
Second, because the class is directly related to local structures
in the spacetime diagram (the phenotype), without the need of some intermediate analysis
such as the calculation of the power spectrum.
This is a prerequisite for the proposed classification approach,
inspired by computer vision.
Third, because the definition of its classes imposes a surjective mapping
from the set of ECA rules to the set of LP classes.
The first two reasons apply to the Wolfram classification as well,
but the third does not: Wolfram classes are qualitative,
and spacetime diagrams of the same rule can be assigned to different Wolfram classes.
This indeterminacy obstructs the generation
of a reliable dataset of rule-diagram pairs,
which is required (or at least strongly preferred)
for properly training a NN (cf.~section~\ref{sec:simple-cnn}).

\begin{table}[t]
    \caption{The five LP classes~\cite{li1990structure} with a brief description
    and the number of rules that each class contains.
    The description applies to the CA dynamics when it has `settled' after some transient period.
    See Fig.~\ref{fig:input-examples} for an example of each class.}
    \begin{tabular}{p{.2\linewidth}L{.67\linewidth}R{.1\linewidth}}
        \hline\noalign{\smallskip}
        \textbf{Class} & \textbf{Description} & \textbf{\# rules} \\
        \noalign{\smallskip}\svhline\noalign{\smallskip}
        Null & Homogeneous fixed-point rules: entirely black or white. & 24\\
        Fixed point & Static or horizontally translating inhomogeneous configurations &  97\\
        Periodic & Recurring configurations with a period of two or three timesteps & 89\\
        Locally chaotic & Chaotic dynamics confined by the domain walls & 10\\
        Chaotic & Rules with random-looking spatiotemporal patterns, or with exponentially divergent cycle lengths as lattice length is increased, or a non-negative spatial response to the perturbations. & 36\\
        \noalign{\smallskip}\hline\noalign{\smallskip}
    \end{tabular}
    \label{tab:Li-Packard-classes}
\end{table}

\subsection{Artificial neural networks for image classification}\label{subsec:ann}

We will perform an automated LP classification by means of a convolutional neural network (CNN). Many excellent monographs and review articles on this popular topic are available; see e.g.~Goodfellow et al.~\cite{goodfellow2016deep} for a general introduction to deep learning, and Dhillon and Verma (2020) for CNNs in particular \cite{dhillon2020convolutional}.
Our choice for CNNs is motivated by four observations that we briefly cover below.\\

First, NNs have long been shown to be capable of ECA classification.
Kunkle~\cite{kunkle2003automatic} has shown (over twenty years ago) that
a very simple fully-connected NN is capable of correctly identifying the LP class
in over $98\%$ of the cases. Kunkle feeds a combination of seven parameter values to the input layer
of the NN. These parameters are based on the rule table (genotype),
rather than on a simulated spacetime diagram.
Whilst our approach will attempt to avoid precisely this kind of a priori knowledge of the CA setup,
the work does present a proof of concept regarding automated classification of ECAs.\\

Second, spacetime diagrams can be interpreted as digital images.
These images are as simple as they come:
single-channel, binary (black-and-white), and with a relatively low resolution.
da Silva et al.~\cite{dasilva2016classification}
(further enhanced by Machicao et al.~\cite{machicao2018cellular})
have shown the feasibility of Wolfram classification of ECAs
by treating the spacetime diagram as a two-dimensional texture. 
They do not use a NN approach, but rather use two conventional texture descriptors:
local binary pattern variance and Fourier descriptors.
Whilst the accuracy for identifying complexity (in Wolfram's interpretation) is only $67\%$,
it is highly accurate for other classes.
Generally, the work of da Silva et al.~demonstrates that observing structures within a spacetime diagram
by treating it as an image containing some texture, is a promising avenue.\\

Third, within the spectrum of artificial intelligence,
CNNs have been shown to be
the most promising tool for image recognition and classification.
Arguably the most notable early publication that has established this claim
is LeCun et al.~\cite{lecun1998gradient}
(for a general introduction to (the terminology of) CNNs,
we refer the reader to this standard work as well).
Images used in that study are typically
not as abstract and `geometrical' as the spacetime diagrams we are considering here,
but the authors make a convincing case for the general applicability of CNNs.\\

Fourth, CNNs have quite recently been used for automated ECA classification
in at least two publications. Silverman~\cite{silverman2019convolutional} constructs
a CNN for automated Wolfram classification. Going one step further,
Comelli et al.~\cite{comelli2021comparing} train a simple CNN for automated class identification
for 11 classifications (all included in Tab.~\ref{tab:martinez-classes}),
effectively comparing how easily the various classifications are taught to this CNN.
Whilst both approaches show that it is indeed feasible to construct a CNN
that is capable of inferring an ECA class simply based on its spacetime diagram,
both also fail to notice that the transition rules hidden within the diagram are a dead giveaway (cf.~section~\ref{subsec:benchmark}).
This implies that the CNN is not necessarily trained to recognise mesoscopic patterns
that human observers would use for classification, but could simply be a (computationally demanding)
way of inferring the local update rule.
This is the issue we will phrase in more detail,
and will attempt to avoid in this chapter, with the aim of improving generalisability.

\subsection{Data and objectives}
\label{subsec:data-and-objectives}

\begin{figure}
    \centering
    \includegraphics[width=\linewidth]{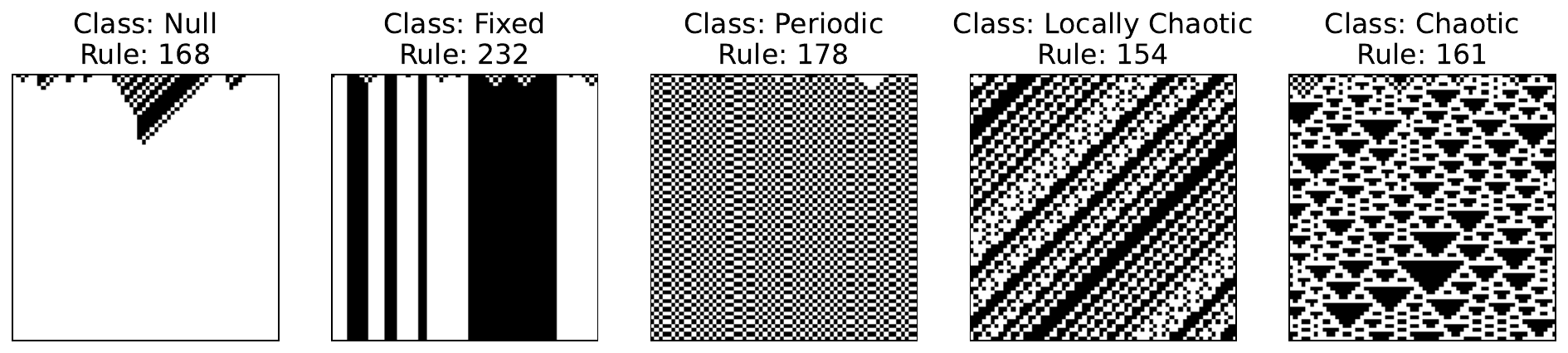}
    \caption{A $64 \times 64$ spacetime diagram for each of the five LP classes, generated from a random initial configuration.}
    \label{fig:input-examples}
\end{figure}

We consider all 256 ECA rules, evolved over 64 time steps, for finite ECAs with $64$ cells and periodic boundary conditions.
This allows for $2^{64}$ initial configurations, resulting in $2^{64} \times 256 \approx 10^{21}$ possible spacetime diagrams.
From this set, we sampled $\num{1024}$ spacetime diagrams for each of the $256$ local update rules for a total of $\num{262144}$ diagrams,
with an additional $128$ diagrams per rule for testing the CNN accuracy (cf.~section~\ref{sec:simple-cnn}).
Fig.~\ref{fig:input-examples} displays an example of these data for every LP class.\\

The objective of the classification task is to predict the LP class, given a spacetime diagram.
We first show that this objective is easily achieved through a grid search (section~\ref{subsec:benchmark}).
We next show that this objective can also be achieved by fitting the diagrams (`data') to the rule or class (`labels') by means of a CNN (section~\ref{subsec:default-CNN}).

\section{A basic CNN for CA classification}\label{sec:simple-cnn}

In this Section we show how the classification task can be solved almost perfectly with a basic algorithm.
We subsequently construct an alternative solution using a CNN and assess its performance.

\subsection{Benchmark: a simple grid search}
\label{subsec:benchmark}

Clearly, simply scanning the spacetime diagram will allow to `fill in' the associated rule table
(all eight `T-tetrominos', such as in Fig.~\ref{fig:rule110_ruletable_diagram}).
In practice, one could simply observe three adjacent cells together with
the central cell in the next time step to fill in the first rule table entry.
Next, shift the observation one cell to the right,
and fill in the next rule table entry -- provided that the ordered triplet of cell states is different.\\

Continuing in this fashion for a random initial configuration,
the probability of encountering all eight triplets
-- and hence completing the rule table -- is over $99\%$
in the first $64$-cell timestep (the first row) alone.
Due to the surjective mapping from rules to LP class,
this simple grid search allows for solving the classification objective
(presented in Section~\ref{subsec:data-and-objectives})
with a near-perfect accuracy in a very efficient way.\\

In some rare cases fewer than all eight rule table entries
are encountered in the $64\times64$ diagram.
Then, for every missing rule table entry,
the probability of still guessing the right rule is halved.
The probability of still guessing the correct \textit{class}, however,
decreases more slowly, as a result of the relationships between rule tables of rules
belonging to the same LP class.\\

Scanning our simulated data set, we find that $370$ diagrams from the training set
and $33$ diagrams from the test set contain incomplete information,
which amounts to $\sim0.1\%$.
Not a single diagram contains fewer than seven rule table entries,
which enables correct LP classification for all but one spacetime diagram,
despite an incomplete rule table.\\

\begin{figure}
    \centering
    \includegraphics[width=\linewidth]{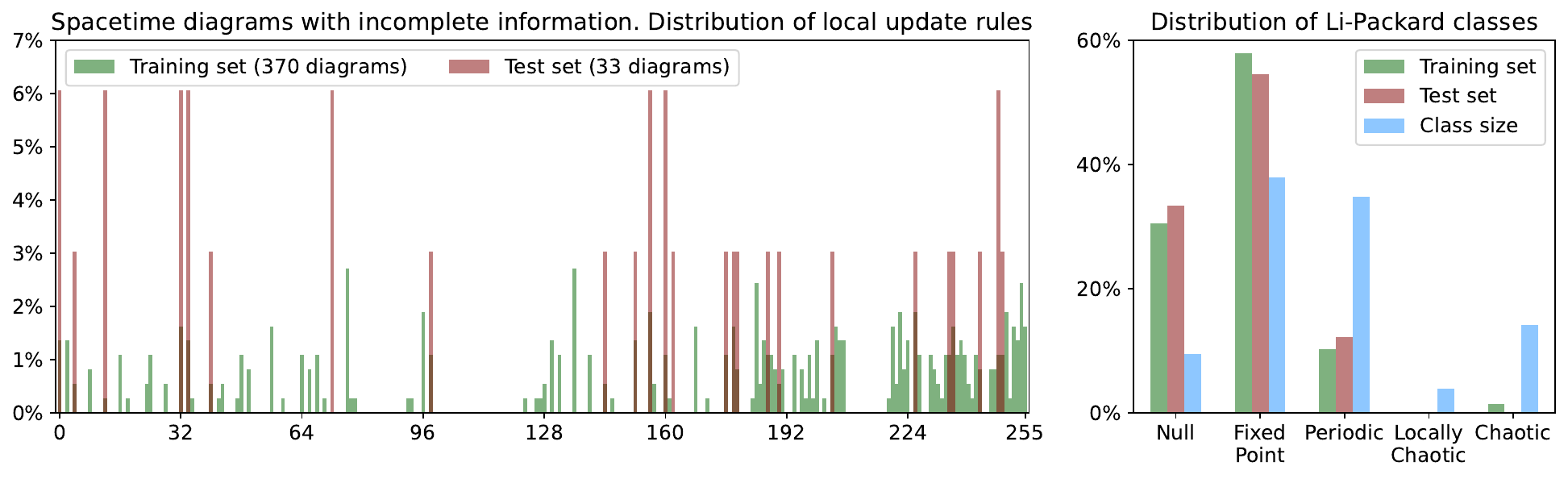}
    \caption{Density histogram of all training set diagrams (green) and test set diagrams (maroon) that contain information that is insufficient for filling in the rule table.
    Left: clustered per elementary rule.
    Right: clustered per LP class that the actual local update rule is associated with, accompanied by a bar denoting the relative sizes of the LP classes (blue).}
    \label{fig:benchmark_undetermined_diagrams}
\end{figure}

\begin{figure}
    \sidecaption
    \includegraphics[width=0.6\linewidth]{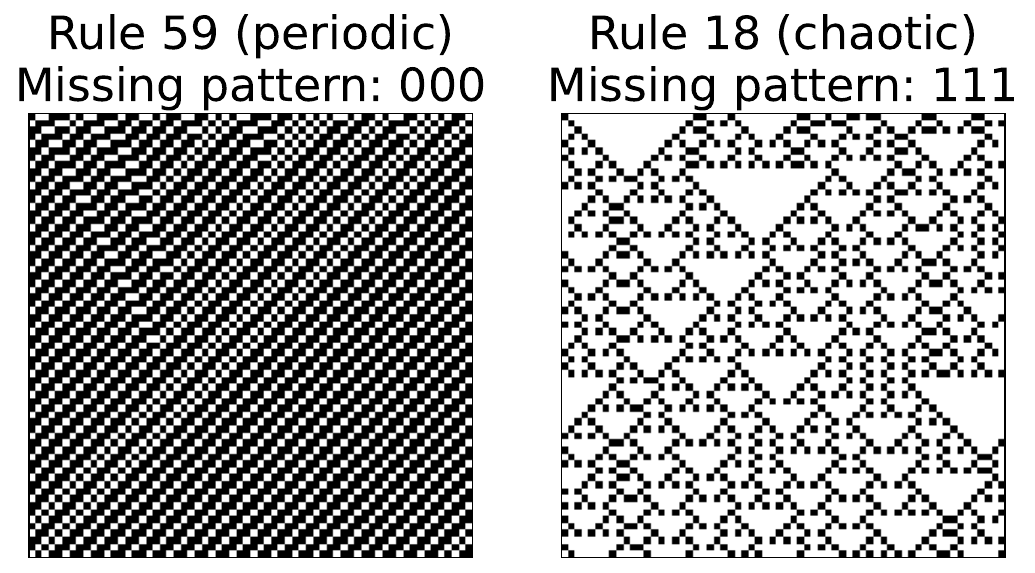}
    \caption{Three remarkable examples of spacetime diagrams that do not contain all rule table entries.
    Left: the only spacetime diagram that could be associated with the wrong LP class following the benchmark algorithm.
    Middle and right: two of the only five of these diagrams that are evolved by a chaotic rule.}
\label{fig:benchmark_undetermined_diagrams_examples}
\end{figure}

Fig.~\ref{fig:benchmark_undetermined_diagrams} displays these results.
Null rules often (and disproportionately) generate missing information,
as do fixed-point rules.
This of course is due to the fact that such rules either annihilate or simply repeat any input information,
respectively, whilst (locally) chaotic rules generate unpredictable patterns.
Additionally, in Fig.~\ref{fig:benchmark_undetermined_diagrams_examples},
two spacetime diagrams with missing information are shown.
These two examples are interesting for different reasons.
On the left, we display the \textit{only} diagram (out of the nearly $3 \times 10^5$ samples)
that could be incorrectly classified (with a probability of $50\%$) using this benchmark scan.
Whilst it is the diagram of a fixed-point ECA,
it could be identified as being in the fixed-point class.
On the right, we show a diagram that is chaotic, but counter-intuitively still manages to withhold an eighth rule table entry.

\subsection{The default CNN}
\label{subsec:default-CNN}

Finite ECA spacetime diagrams can be interpreted as digital images,
which allows our objective to be interpreted as a well-defined computer vision task.
Given the massive data set at hand, it should be possible
to train a CNN to correctly infer the class of a diagram it has not encountered before.\\

The possibility of this approach has been explored by Silverman \cite{silverman2019convolutional}
for the Wolfram classification on ECAs,
reporting an accuracy of over $99.7\%$ whilst training,
and a perfect $100\%$ score on the test set.
More recently, Comelli et al.~\cite{comelli2021comparing}
used a similar simple CNN to compare 11 variations of ECA classification,
reporting an accuracy of $97.58\%$ for LP classification.
These extremely high accuracies can be understood by acknowledging that
for many ECA classifications, we are feeding a solvable problem to the algorithm
-- as was demonstrated in the benchmark grid search (section~\ref{subsec:benchmark}).\\

\begin{figure}
    \centering
    \includegraphics[width=\linewidth]{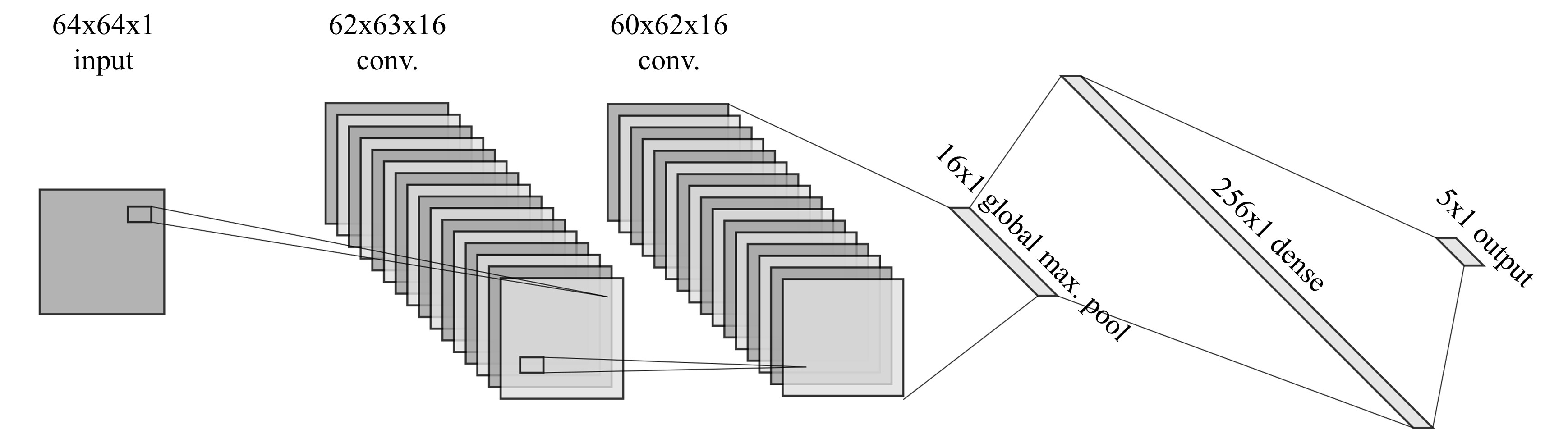}
    \caption{Visual overview of the CNN largely reproduced from Silverman~\cite{silverman2019convolutional},
    essentially chaining two 16-channel convolutional layers,
    a max pooling layer,
    a 256-node dense layer,
    and a 5-node output layer.
    The convolutional kernel has dimensions $3\times2$.
    The model for inferring the local update rule omits the final layer.
    This default CNN for rule inference and LP class inference has a total
    of $\num{6016}$ and $\num{7301}$ trainable parameters, respectively.}
    \label{fig:default_nn_architecture_NN-SVG}
\end{figure}

As a first step, we reproduce the CNN architecture proposed by
Silverman~\cite{silverman2019convolutional,fernandes2019nnclassification},
briefly explaining every design choice
(see Fig.~\ref{fig:default_nn_architecture_NN-SVG} for a visual overview).

\subsubsection{Architecture and activation functions}
\label{subsubsec:default-CNN}
From input to output, the information stream encounters two consecutive convolutional layers,
a global maximum pooling layer, and a $256$-node dense layer.\\

\begin{figure}
    \centering
    \includegraphics[width=\linewidth]{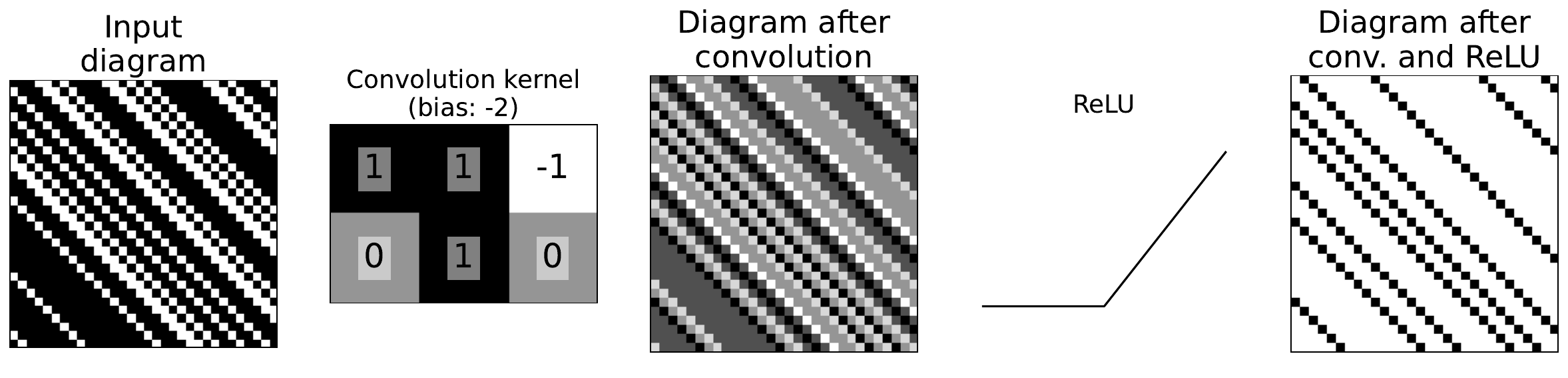}
    \caption{Example of the information flow (left to right)
    whilst consecutively performing a convolution and executing a ReLU activation function.
    This particular choice of convolutional kernel filters out all locations
    that have a $(1,1,0) \mapsto 1$ local update mapping,
    ignoring the state of the resulting cell's neighbours.}
    \label{fig:example-convolution-and-relu}
\end{figure}

The input layer is a single-channel black-and-white image with a resolution of $64\times64$.
Next is a $16$-channel convolutional layer,
where every channel contains the result of a $3\times2$ convolution
for a particular choice of values for the six weights and one bias parameter.
This sums to $112$ free parameters.
By choosing $16$ channels,
the weights and biases for each channel can be optimised to recognise a T-tetromino from the rule table,
that comes in $16$ variations.
Due to the shape of the convolutional kernel,
the leftmost and rightmost columns of pixels cannot be convolved,
and neither can the bottom row.
The resulting channels therefore have a resolution of $62\times63$.
The output of this convolution is activated by means of a rectified linear unit (ReLU),
simply defined as $x \mapsto \text{max}(0,x)$.
For an intuitive understanding of the information flow, see
Fig.~\ref{fig:example-convolution-and-relu}.
A similar convolutional layer maps to $16$ $60\times62$ channels, again ReLU-activated,
for further increasing the spatial sensitivity and overall power of the model.
This layer contributes $\num{1552}$ trainable parameters.\\

In the information flow we next have a bottleneck:
the global max pooling layer simply keeps only the maximum value
of each of the $16$ channels, in order to filter out noise.
After the bottleneck comes a fully-connected layer of $256$ nodes,
i.e.~each of the maximum values of the $16$ convolutional channels
is connected to all of the $256$ nodes, for a total of $4096$ weights and $256$ biases.
Note that this layer contributes over half of the total number of trainable parameters.
If the objective is to predict the ECA rule, the $256$-node layer is the output layer.
If, however, we aim to predict the LP class,
the values in the previous layer are connected to a five-edge output layer.
The values from the final layer
are mapped to the unit interval by means of a SoftMax function (similar to the hyperbolic tangent function),
such that the output vector can be read as a discrete probability distribution.

\subsubsection{Training strategy}

With the architecture outlined in Section~\ref{subsubsec:default-CNN} in place, the $\num{6016}$ or $\num{7301}$ free parameters
need to be inferred.
This is achieved by back-propagation using the highly convenient Keras infrastructure~\cite{chollet2015keras}.\\

In short, first the generated data set of $2^{18}$ diagrams is randomly split
in $3/4$ training set and $1/4$ validation set.
No cross-validation is required due to the size of the data set.
Next, labels associated with each diagram are translated to one-hot vectors
and compared with the CNN's outcome by means of categorical cross-entropy.\\

We select a batch size of $64$: a new set of parameter values is proposed after evaluating $64$ diagrams.
This effectively means that the back-propagation method implements $\num{3072}$ parameter updates for each time the full training set is evaluated (each `epoch').
Weights and biases are initialised randomly,
and updated by means of Keras's built-in Adam optimiser.
The optimiser's learning rate parameter
-- which determines the velocity at which the optimisation process moves through parameter space --
is set at $10^{-3}$.
We run over maximally $50$ epochs, but abort training when the difference in accuracy on the validation set is less
than $10^{-5}$ over $5$ subsequent epochs.\\

The model is trained using a NVIDIA T4 GPU,
running over a single epoch in approximately $25$ seconds.

\subsection{Performance assessment of the default model}

The default model achieves an accuracy of $99.99\%$ for both the training set and the validation set for the LP classification,
stopping its training after $12$ epochs. From the $\num{65536}$ diagrams in the validation set, the model infers the wrong LP class for only six.
Note -- whilst this was not our objective -- that the accuracy we achieved
is much higher than the one reported by Comelli et al.~\cite{comelli2021comparing},
which may be partly due to our larger training set.
Our training accuracy is even higher than the one reported by Silverman~\cite{silverman2019convolutional}. Clearly, his reported $100\%$ for test set accuracy is not surpassed, but to be fair (looking at the benchmark results) this perfect score is probably rather lucky.
Keep in mind, moreover, that we have chosen a LP-based classification rather than Wolfram's.
For rule identification rather than LP classification,
the model achieves an accuracy of $99.86\%$ on the training set,
and $99.85\%$ on the validation set.
In both cases, the accuracy history exceeds $99.5\%$ after the first epoch,
displaying the power given to the model via the large number of free parameters.\\

\begin{figure}
    \centering
    \includegraphics[width=\linewidth]{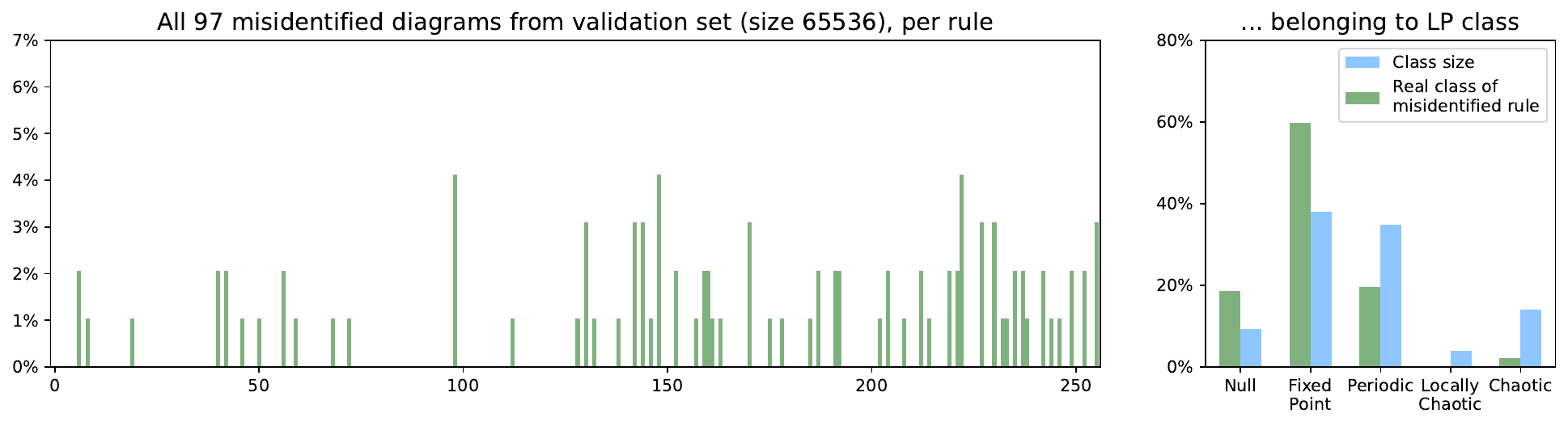}
    \caption{The default CNN achieves a validation accuracy of $99.84\%$
    when inferring the local update rule from the spacetime diagram.
    Most incorrectly identified rules belong to the periodic LP class.
    None belong to the locally chaotic class.}
    \label{fig:default-model_misidentified-val-set}
\end{figure}

\begin{figure}
    \centering
    \includegraphics[width=\linewidth]{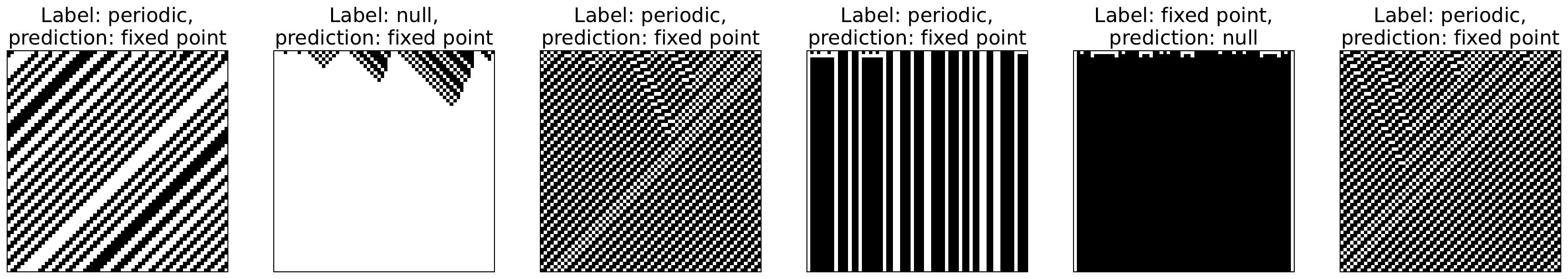}
    \caption{The default CNN achieves near-perfect accuracy on the validation set
    when directly classifying the diagrams into one of five LP classes.
    Here we show all six wrongly classified diagrams.
    Note that these errors are easily forgiven as borderline examples.}
    \label{fig:default-model_all-wrong-LP-classes}
\end{figure}

Results for rule identification of the validation set are shown in Fig.~\ref{fig:default-model_misidentified-val-set}.
Comparing these results to Fig.~\ref{fig:benchmark_undetermined_diagrams},
we observe largely the same trends.
Due to the near-perfect accuracy, however, the standard deviation of the wrongly-classified diagrams is quite high
and results are to be compared with some leniency.
Fig.~\ref{fig:default-model_all-wrong-LP-classes} shows all six wrongly classified diagrams in the LP classification.\\

The goal, now, is first to trim down the clearly overly powerful model.
Next, we experiment with a number of changes to the data and the neural architecture,
exploiting characteristics of the classification objective.

\section{Variations and extensions of the classification CNN}\label{sec:extensions-limits}

In this Section we first design an algorithm that is equally good at determining the ECA rule (and hence the LP class) as the benchmark grid search,
whilst following the architecture of a CNN.
Next, we show how the CNN of Section~\ref{subsec:default-CNN}
can be re-designed and simplified with the objective of class identification
in mind, rather than rule identification.
Next, we explore various methods of data augmentation that again
reflect this research objective.
Third, we assess and visualise the performance of the
maximally-simplified model with appropriate data augmentation.

\subsection{A perfect CNN for rule identification}

\begin{figure}
    \centering
    \includegraphics[width=\linewidth]{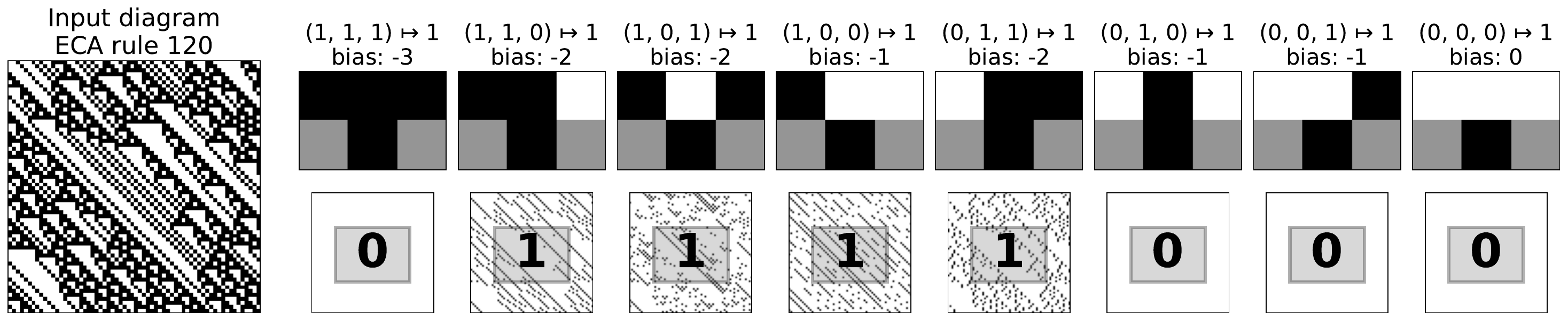}
    \caption{By subsequently applying a convolution and a ReLU activation,
    any ECA diagram can be decomposed into eight channels.
    The channel contains only $0$s if the particular neighbourhood outputs $0$,
    and contains $1$s if this neighbourhood outputs $1$.
    The location of the $1$s in the convolutional channel correspond with the location of the T-tetromino in the input diagram.
    Whilst this has a certain aesthetic quality to it,
    the only aspect that matters to our CNN is the maximum value over the entire convolutional channel,
    considering that each channel is subsequently subjected to a global maximum pooling. The resulting octuple is the binary representation of the local update rule.}
    \label{fig:convolutional-decomposition-example_120}
\end{figure}

Any ECA diagram that contains all neighbourhood configurations
can be decomposed into its `neighbourhood contributions'
via a convolution into eight channels.
An example for a single channel was shown in Fig.~\ref{fig:example-convolution-and-relu},
and Fig.~\ref{fig:convolutional-decomposition-example_120} contains a full decomposition for a diagram of rule $120$ ($01111000_2$ in binary).\\

The global maximum of these channels is either $0$ or $1$,
and the ordered octuple of maxima encodes the local update rule.
We feed this binary pattern into a $256$-node output layer,
once more by manually choosing the weights of each node $i$:
\begin{align*}
    \mathbf{w}_i &= (w_{i,0}, \dots, w_{i,j}, \dots, w_{i,7}),
    \quad \text{with } w_{i,j} = 2\, \text{bin}(i)_j - 1,
\end{align*}
where $\text{bin}(i)_j$ is the $j$th binary digit of the integer $i$.
The corresponding biases are
\begin{align*}
    b_i &= 1 - \sum_{j=0}^{7} \text{bin}(i)_j.
\end{align*}

The output layer is again mapped to a probability with a SoftMax function.
The resulting CNN has $2360$ (fixed) parameters.
It incorrectly identifies only $331$ diagrams in the training set, and as such reaches an accuracy of $99.87\%$ for rule identification. This former number is close to the number of  diagrams (370) that contain an insufficient amount of information for a decisive grid search.\\

Clearly, this approach for retrieving the local update rule is far more tedious than simply scanning the grid. So, if we are to harvest the power of CNNs,
we would do better to design them for actual pattern recognition,
rather than use them as a long computational detour for an essentially trivial task.

\subsection{Trimming the default CNN for mesoscopic pattern recognition}
\label{subsec:trimmed-cnn}

We trim down the default CNN architecture for two reasons.
First, it demonstrates that even simpler models are capable of reaching a high classification accuracy,
even if they are not tailored for reconstructing the rule table.
Second, using these trimmed-down versions of the CNN, the accuracy decreases slightly,
which provides the headroom that is required to inspect the effect
of the various data augmentations that we will explore in Section~\ref{subsec:data-augmentation}.\\

When re-designing the NN, a first obvious observation
is that we must remain true to the convolutional approach,
cleverly exploiting the local structure.
For comparison: for our dataset,
a fully-connected NN with a single $256$-node hidden layer performs acceptably,
with a validation accuracy for LP classification of just over $95\%$.
However, it requires well over a million trainable parameters.
We therefore explore variations on the default CNN presented in Section~\ref{subsec:default-CNN}.\\

Guided by the objective of designing a very simple CNN that is good at predicting
LP classes, we considered a number of architectures
that, next to the general principles of successful CNN design \cite{dhillon2020convolutional}, all have the following three additional principles in common.
First, the convolutional kernel of the first hidden layer must not envelope
an ECA neighbourhood and the resulting subsequent state (a T-tetromino).
Second, the CNN should have a sufficiently large receptive field,
enabling the observation of mesoscopic structure.
Third, the hidden dense layer before the five-fold output layer must not
be capable of encoding the ECA rule.
Note that none of these principles are explicitly reflected
in the default CNN of Section~\ref{subsec:default-CNN}.\\

\begin{figure}
    \centering
    \includegraphics[width=\linewidth]{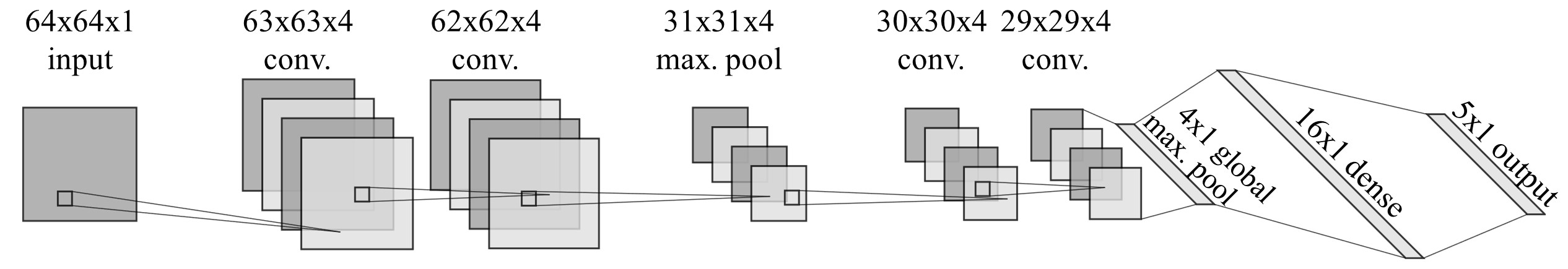}
    \caption{Architecture diagram of the CNN that has been trimmed
    according to the principles laid out in Section~\ref{subsec:trimmed-cnn}.
    The convolutional kernels (not to scale) have dimensions $2\times2$.
    The 16-node hidden dense layer and the 5-node output layer is changed
    to a single 256-node output layer when training for rule identification.
    This trimmed-down CNN only has 389 (resp.~$\num{1504}$) trainable parameters
    for LP class (resp.~rule) classification.} 
    \label{fig:trimmed-CNN-architecture}
\end{figure}

The CNN shown in Fig.~\ref{fig:trimmed-CNN-architecture} fulfils the above principles. The convolutional kernel has dimensions $2\times2$, which cannot fit a T-tetromino.
The two additional hidden convolutional layers and (especially) the central maximum-pooling layer increase the receptive field substantially.
The final hidden layer is a 16-node dense layer, which certainly cannot uniquely encode information on ECA rules.
A validation accuracy of at least $99.10\%$ can be achieved with this CNN 
for LP class identification, whilst it comes with only $389$ trainable parameters.
For rule identification, however, the validation score drops
substantially, to $71.19\%$, despite involving $1504$ parameters.

\subsection{Data augmentation tailored for class identification}
\label{subsec:data-augmentation}

We want to train the model Fig.~\ref{fig:trimmed-CNN-architecture} such that it is sensitive to what matters
to the classification, and insensitive to what does not.
One way of doing so is by manipulating the data set in a way
that affects the content at pixel level, but does not affect
the overall pattern structure of the spacetime diagram.
In other words, to alter it in a way that does not make
it too difficult for a human to still perform the classification.
This is one aspect of a type of regularisation known as data augmentation~\cite{shorten2019augmentation}.
Data augmentation is a preprocessing technique that is typically called upon
to artificially enlarge and diversify the size of the dataset,
such that the CNN is not overfit on the (often sparse) training data.
Due to the practically unlimited size and maximal diversity of the set of ECA spacetime diagrams, overfitting is not an issue, though we will use data augmentation techniques to average non-essential aspects of the data,
as such aiding the algorithm to focus on what we want it to.
Importantly, this will not improve validation accuracy, but it will increase the gap between the accuracies of predicting the rule versus the LP class.\\

\begin{figure}
    \centering
    \includegraphics[width=\linewidth]{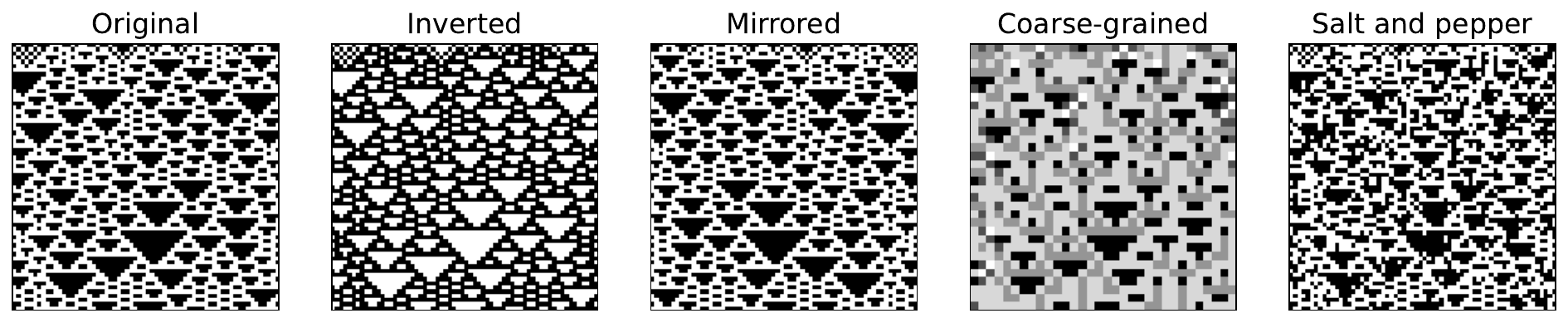}
    \caption{An example of a spacetime diagram in the LP class `chaotic' (left),
    with the effect of the four considered data augmentations displayed
    to the right. In these images, coarse-graining was performed in $2\times2$ windows,
    and we added $5\%$ salt-and-pepper noise.}
    \label{fig:augmentation-examples}
\end{figure}

Fig.~\ref{fig:augmentation-examples} visualises the four types of data augmentations that we tested:
inversion, mirroring, coarse-graining, and adding salt-and-pepper noise.
First, the way one decides to colour the diagram is not of any
significance to the underlying binary structure, so inverting
this colour should not influence the LP classification.
Second, neither should horizontally flipping because the emergent patterns are not fundamentally different when they manifest from left to right or vice versa.
The mirror and inversion operations (and their combination) are of course precisely what define the subsets of ECA rules that are equivalent to each other,
such that only $88$ `independent' rules remain~\cite{li1990structure}.
Third, the patterns and mesoscopic structures in the spacetime diagrams should still be distinguishable when they are slightly blurred. In particular, we can coarse-grain the image such that, in blocks of $2\times2$, every pixel value is changed to the average value in that block.
Clearly this removes some information on the microscopic level, disrupting rule identification.
Fourth, we explore the addition of so-called salt-and-pepper noise,
which simply means that a certain percentage of pixel values are randomly changed to 0 or 1. This will introduce contradictions in the  rule table, whilst mostly leaving the larger structures untouched.\\

In practice, data augmentation is performed while training the CNN, i.e. `online'. Every batch of $64$ diagrams
is randomly augmented using one or more of the augmentation techniques.
A diagram has a probability of $N/(N+1)$ of being augmented by one (and only one) of $N$ selected augmentations, and a probability of $1/(N+1)$ of remaining unaffected.\\

Additionally, we take into account the equivalence between rules in the training phase as well.
We evaluate the CNN's capacity of determining which of the $88$ equivalence classes the rule belongs to (cf.~Tab.~1 in~\cite{li1990structure}),
by decreasing the number of nodes in the final layer from $256$ to $88$ and adapting the diagram labels accordingly.
The resulting CNN contains $664$ trainable parameters.

\subsection{Performance assessment of the trimmed CNN on augmented data}

We trained the CNN with different combinations of data augmentation techniques,
with various learning rates, and with noise levels varying between $1\%$ and $10\%$.
The highest accuracies achieved on the validation set 
are displayed in Tab.~\ref{tab:assessment_data-augmentation} for no augmentation, individual augmentation techniques, and a combination of all techniques.\\

\begin{table}
    \caption{Performance assessment of the trimmed CNN (Fig.~\ref{fig:trimmed-CNN-architecture})
    with various forms of data augmentation.
    We list the highest attained accuracies on the validation set. We distinguish the identification of the actual local update rule (left), the independent rule (middle), and the LP class (right). Data augmentation negatively impacts all accuracies, but (crucially) simultaneously increases the performance gap between rule and class identification.} 
    \begin{tabular}{p{3.5cm}L{2.3cm}L{3.2cm}L{2.2cm}}
    \hline\noalign{\smallskip}
    Augmentation technique & All rules ($256$) & Independent rules ($88$) & LP classes ($5$) \\
    \noalign{\smallskip}\svhline\noalign{\smallskip}
    None & $71.19\%$ & $80.58\%$ & $99.10\%$ \\
    \noalign{\smallskip}\svhline\noalign{\smallskip}
    Inverted & $28.33\%$ & $81.73\%$ & $98.31\%$ \\ 
    Mirrored & $48.85\%$ & $78.81\%$ & $98.66\%$ \\
    Coarse-grained ($2\times2$) & $15.75\%$ &$65.35\%$ & $97.83\%$ \\
    Salt-and-pepper noise ($1\%$) & $66.54\%$ & $75.52\%$ & $97.18\%$ \\
    \noalign{\smallskip}\svhline\noalign{\smallskip}
    All augmentations & $28.25\%$ & $56.17\%$ & $95.42\%$ \\
    \noalign{\smallskip}\hline\noalign{\smallskip}
    \end{tabular}
    \label{tab:assessment_data-augmentation}
    \end{table}

As expected, all accuracies decrease compared to the CNN trained on the non-augmented data.
While the LP classification accuracy remains above $97\%$,
the rule identification accuracy decreased significantly,
and -- interestingly -- differently for the various augmentation techniques.\\

When inverting or mirroring the diagrams, it becomes much more difficult
to detect the actual rule, because the augmentation effectively transforms the local update rule into a different (equivalent) one. The effect is larger for inversion, which might be caused by convergence difficulties while training the CNN.
However, this is probably mostly due to the fact that $64$ rules are their own mirror image, whilst only $16$ rules are their own inversion.
That is to say: the mirror operation often is the same as the identity operator on the microscopic level,
effectively nullifying the augmentation altogether.
Moreover, the relation between inversion and mirror augmentation on the one hand,
and equivalence classes of local update rules on the other,
is clear from the results as well:
the accuracies on independent rule identification are largely unaffected
(or even go up) when applying these data augmentations. \\

Coarse-graining by averaging every $2\times2$ block of pixels
clearly has the strongest effect on the CNN's capacity for rule detection,
bringing down the accuracy from over $71\%$ to less than one in six.
Whilst the number of wrong LP classifications more than doubled,
the gap with rule detection is by far the largest of the considered augmentation techniques.\\

We observed that even a little noise was enough to negatively affect
all accuracies, without favouring LP classification, which is undesirable in terms of our objective.
The values reported in Tab.~\ref{tab:assessment_data-augmentation} correspond to
a noise level of only $1\%$.
Clearly, adding this type of noise does not impact the CNN's ability to
achieve our objectives in a positive way.\\

In order to inspect any kind of interference between different augmentation techniques, we also investigated their combined effects.
This results in accuracies that are significantly smaller than the average of all accuracies of single-technique augmentations,
which is testimony to the fact that convergence to an optimum is harder to achieve
when the training process is `distracted' with augmentations.\\

\begin{figure}
    \sidecaption
    \includegraphics[width=.55\linewidth]{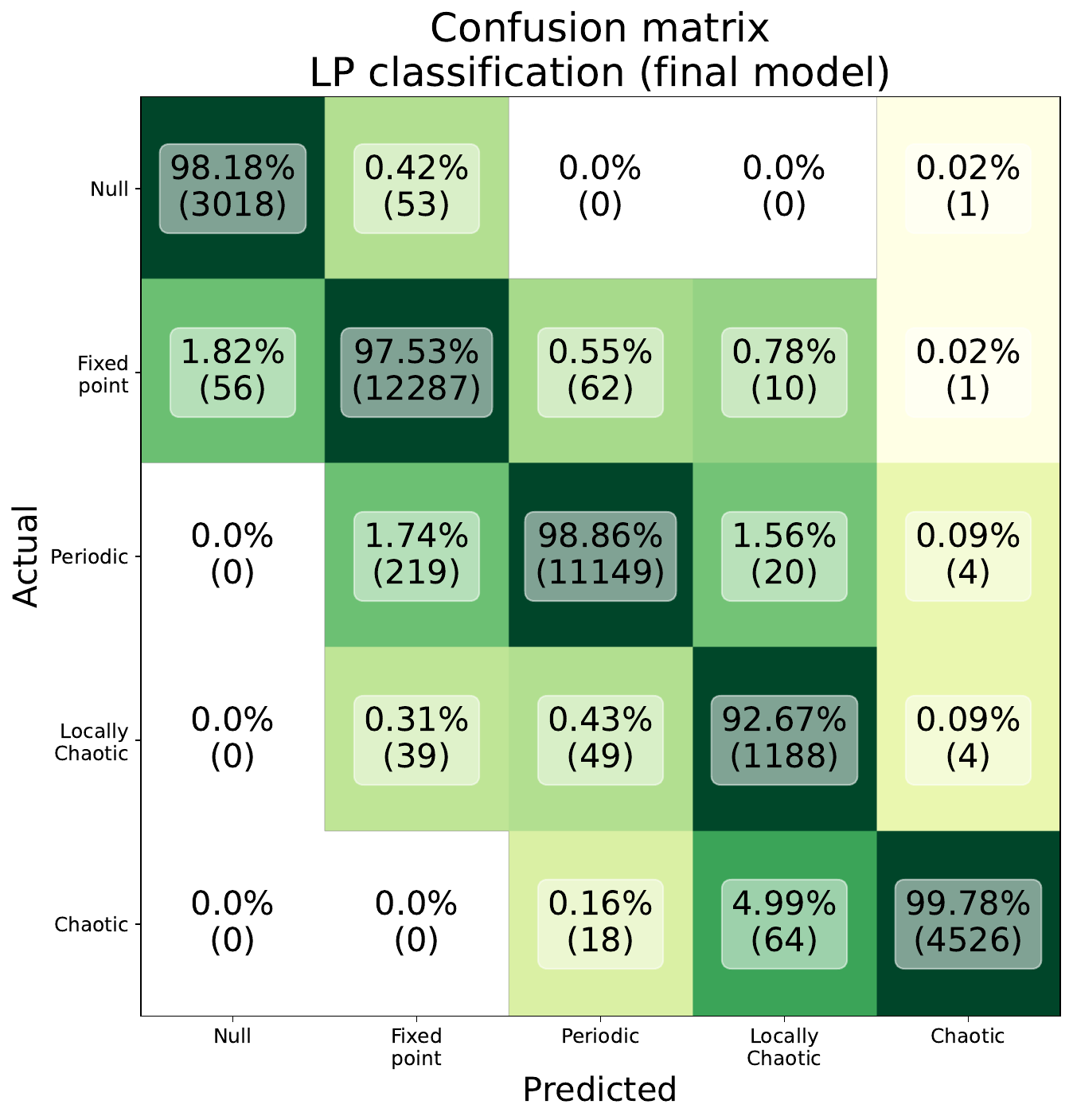}
    \caption{The confusion matrix for LP classification by our final CNN model (Fig.~\ref{fig:trimmed-CNN-architecture}),
    trained with all augmentation techniques except salt-and-pepper noise,
    normalised over the predictions.
    This model reaches an accuracy of $98.17\%$ on the test set ($128$ diagrams per ECA rule),
    compared to $61.25\%$ for identifying one of the $88$ independent rules.
    Relatively often (in almost $5\%$ of the cases) the model wrongly identifies chaotic diagrams as locally chaotic.}
    \label{fig:confusion-matrix-final}
\end{figure}

Informed by the results in Tab.~\ref{tab:assessment_data-augmentation},
we perform a final optimisation on the complete LP-classified training set
(including the validation set), with all augmentations except salt-and-pepper noise.
This yields an accuracy on the test set of $98.17\%$ for the LP classification,
and $61.25\%$ for identifying the independent rule.
Fig.~\ref{fig:confusion-matrix-final} shows the confusion matrix for the LP classification,
summarising which predictions were made by the CNN.

\section{Discussion and future developments}

The overlap between the research domains of deep learning on the one hand,
and discrete dynamics systems on the other, continues to grow.
In the case of CAs, this is due to the readily available vast computational resources
and to exciting developments in the study of computer vision.
Due to their simple, convolution-like nature, CAs have been used
for studying the internal mechanism of CNNs~\cite{gilpin2019convolutional}.
In contrast, here we examined techniques where CNNs can aid CA research.
Such techniques -- especially cleverly designed CNNs -- can help to identify
a CA spacetime diagram as belonging to a particular behavioural class.
Automated classification can therefore be a very useful tool for
identifying large numbers of spacetime diagrams.
In turn, this is an important requirement to make large-scale statistical
assessments of classification schemes practically feasible,
which would help formulate an answer to the first original problem in the theory of CAs~\cite{wolfram1984twenty}.\\

This has to be done right, however. In this chapter, we first explicitly showed how easy it is to find a perfect automated
classification of ECAs when `cheating' is allowed, i.e.~by simply scanning the local update rule or by designing a CNN and hand-picking the weights and biases.
Next, we started from existing CNN implementations for automated classification, but re-designed them.
We did so in such a way that the algorithm picks up on the mesoscopic structures
that determine their phenomenology -- and hence their classification --
whilst having a hard time identifying the microscopic structures that directly reveal their genotype.
This redesign was guided by first altering the architecture itself,
making it less obvious for the information flow to contain a direct encoding of the local update rule.
Second, we considered four techniques of data augmentation,
and demonstrated that out of these four, adding coarse-graining is the best way to keep
the model from learning the underlying rule.
Our final model was trained using all augmentations except salt-and-pepper noise.
LP classes were assigned erroneously in less than $2\%$ of the space-time diagrams.
An important note for future development is that this CNN relatively often wrongly identifies chaotic or periodic diagrams as locally chaotic.\\

The future of the research on the edge between CAs and deep learning is first of all to explore more architecture variations and training strategies,
to increase the gap between rule accuracy and class accuracy.
Second, we may explore different classification schemes,
much like Comelli et al.~\cite{comelli2021comparing} did,
but now again with the goal of finding the best way to maximise the performance gap.
After all, it is not obvious that the LP classification is the most useful,
especially not when it comes down to pinpointing Wolfram's `class-IV' complex behaviour~\cite{li1990structure}.
Next, armed with a good automatic classification scheme that is sensitive to mesoscopic patterns, the CNN may be used for classification of non-elementary CAs,
such as CAs with a larger neighbourhood or non-uniform CAs.\\

One must remain critical when interpreting the inferences obtained using  a deep learning model, especially when it is used on data that is was not trained on.
On the other hand, this is arguably the best we can do for a supervised learning approach without spending ages of mind-numbing and (at least partially) subjective manual labelling
for the creation of a proper training set.
Because of that reason, another promising approach in automated classification
is self-supervised learning~\cite{rani2023selfsupervised}, where the algorithm decides `for itself'
which phenomenologies are to be grouped together, without the need for a label.
This approach will be the topic of our forthcoming work,
and is a powerful ally to the supervised approach presented in this chapter,
as it may uncover different aspects of the same problem.\\

Looking further, many research opportunities on the edge of CAs and NNs are still open for discovery.
One such possibility is the mobilisation of time series in the context of AI-aided CA classification.
This implies deriving a time series from the CA evolution and using, for example, the time-dependent spatial entropy as an input for a recurrent neural network \cite{goodfellow2016deep}.
Whilst this kind of an approach does not strictly contain more information
than the full spacetime diagram, it may help to direct the algorithm's focus,
much like we did for CNNs in this chapter.
Yet another next step would be not to predict classes,
but to predict values associated with the CA that are otherwise computationally demanding
to calculate directly, such as the Lyapunov spectrum (cf.~M.~Vispoel's contribution in this book).
A further exciting possibility is the implementation of (modest) generative AIs,
where a particular local update rule is selected that generates
a desired emergent behaviour (see e.g.~Mordvintsev et al.~\cite{mordvintsev2020growing}),
effectively turning the classification problem on its head.\\

In one way or another, all these roads lead to the identification of links
between interesting emergent behaviour on the one hand
and, on the other hand, the very simple algorithmic sequences at the heart of CAs.
This further strengthens both our mathematical understanding of this fascinating dynamical model,
and further enables applications in computer science and mathematical modelling.


\end{document}